\documentclass[final]{aipproc}
\layoutstyle{6x9}
\newcommand\doingARLO[2][]{%
  \ifx\mmref\undefined #1\else #2\fi
}
  
\begin{document}

\title{Experimental Observation of Coherence and Stochastic Resonances in an Electronic Chua Circuit}

\author{Oscar Calvo$^{*}$, Iacyel Gomes$^{*}$, Claudio R. Mirasso$^{*}$ and Ra\'{u}l Toral$^{*,\dagger}$}
{address={$^*$ Departament de F\'{\i}sica, Universitat de les Illes Balears,  07071-Palma de Mallorca, Spain\\ 
$^{\dagger}$ Instituto Mediterr\'aneo de Estudios Avanzados ,
CSIC-UIB, 07071-Palma de Mallorca, Spain
}}

\begin{abstract}
Stochastic and coherence resonances appear in nonlinear systems subjected to an external source of noise and are characterized by a maximum response at the optimal value of the noise intensity. This paper shows experimentally that it is possible to observe them in a {\sl chaotic} system. To this end we have analysed an electronic Chua circuit running in the chaotic regime and added noise to its dynamics. In the case of coherence resonance, we observe an optimal periodicity for the jumps between chaotic attractors, whereas in the case of stochastic resonance we observe a maximum in the signal--to--noise ratio at the frequency of an external sinusoidal perturbation. 
\end{abstract}
\maketitle

\section{Introduction}
Contrary to the intuitive role of noise as a disordering agent, it
has been shown that the response of a nonlinear dynamical system can
improve in some occasions by the addition of noise. The
paradigmatic and pioneering example is that of {\sl stochastic
resonance} in bistable systems\cite{BSV81,NN81} which had been
originally proposed as a mechanism for the observed periodicity of
the Earth's ice ages. In this case, a dynamical system with two
stable fixed points is subjected to the action of a periodic forcing
and noise. The periodic forcing is sufficiently weak that, by itself,
can not overcome the separation barrier between the stable fixed
points and the system oscillates around one of them, chosen by the
initial conditions. The presence of noise, however, allows the system
to make random excursions and visit both fixed points. The jumps
between the fixed points are easier to make when the effect of the
external forcing is to lower enough the separation barrier. If the
average time for the stochastic walk to go from one fixed point to
the other coincides with the periodicity of the external signal, a
maximum coincidence of the jumps with the external period is
observed. The coincidence is optimal for the right value of the noise
intensity. This is, basically, the mechanism for stochastic
resonance. The original example has been extended to many other cases
of physical as well as biological interest and there exists now a
well established number of results concerning theoretical, numerical
and experimental studies of stochastic resonance under a large
variety of conditions\cite{JSP70,GHJM98,ANMS99}.

A related but different phenomenon is that of {\sl coherence
resonance} by which a quasi-periodic signal is generated under the
effect of the fluctuations in a system not subjected to a periodic
forcing\cite{GDNH93,RS94,PK97}. Again, the periodicity is optimal
(resonance) for a given value of the noise intensity. The simplest
example is an excitable system. This is a system with a fixed point
which is stable for low amplitude perturbations. However,
perturbations larger than a threshold value result in the system
returning to the fixed point after a large excursion, a pulse, in
phase space. In the case of coherence resonance, the perturbations
are induced by noise and, if the noise has the appropriate intensity,
the system is excited at the end of every pulse in such a way that a
quasi-periodic train of pulses is generated. Coherence resonance has
also been observed in bistable and oscillatory
systems\cite{LS00,POC99} or close to the onset of a
bifurcation\cite{NSS97}. Experimental evidence has been shown for
excitable electronic circuits\cite{PHYS99,HYPS99} and a laser
system\cite{GGBT00}.

In this paper we show experimental evidence for both resonances in
electronic Chua circuits operating in the chaotic regime. The
possibility of having stochastic resonance in a chaotic circuit with
two attractors under the presence of noise had been shown
theoretically and numerically by Anishchenko and
coworkers\cite{ASC92,ANMS99}. This is different from the so--called
{\sl deterministic stochastic resonance}\cite{ANS93} which appears in
deterministic chaotic systems close to the onset of intermittency and
that has been shown experimentally in ferromagnetic\cite{RJBB97} and
laser\cite{PC98,CVC00} systems. 

\begin{figure}[htb]
\rotatebox{-90}{\includegraphics[width=105mm,height=170mm]{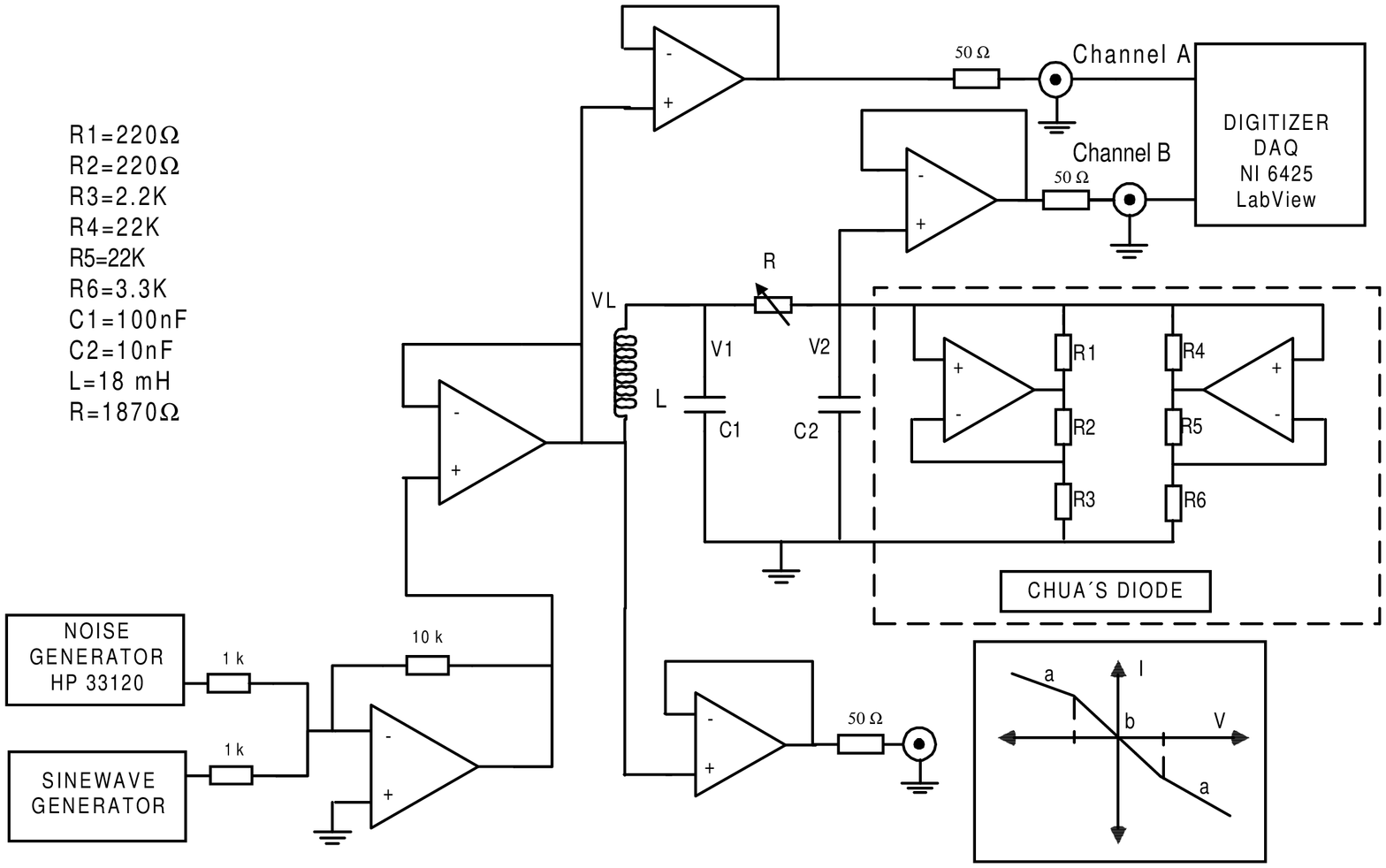}}
\end{figure}
\vspace{-3.0cm}

\begin{center} 
{\bf FIGURE 1} Experimental setup.
\end{center}

Our Chua circuit is shown in figure 1. A digital acquisition
board from National Instruments NI-DAQ plugged into a laptop computer was used to digitize the signal at a sampling rate of $10$KHz,. A Labview program
controlled the board in a continuous acquisition mode and an HP 33120 function
generator was used to provide the noise signals from $0$ to $5$ V[rms].
A Chua circuit can be described in terms of three non-linear first
order differential equations\cite{chua}. These equations predict that the system
has, for some range of parameters, three unstable fixed points. In
this case, the output voltages $V_1$ and $V_2$ oscillate chaotically
in time around different dynamical attractors. In our experiments, we
choose  the main parameters $C_1=10$ nF, $C_2=100$ nF, $R=1673$ $\Omega$, $a=-8/7$,
$b=-5/7$, where $a$ and $b$ characterize the slope of the non-linear
Chua resistance (see inset fig. 1). For this set of
parameters, there is a {\sl single scroll} attractor (as shown in the
fig. 2(a), and its mirror image (not shown in the figure). In the
absence of any external perturbation, $V_1$ and $V_2$ chaotically
oscillate around one of the unstable fixed points with a mean
oscillation frequency of $\sim$ $2.3$ KHz and without any possibility
to jump to its mirror image attractor.  Depending on the
initial conditions, the system moves within this attractor or on its
mirror image. The time evolution of $V_2$ is
shown in the figure 2(b). We can see that $V_2$ (and similarly for $V_1$)
oscillates chaotically with a rather large fluctuations in the minima
and maxima values. This fact will be important when studying the
stochastic and coherence resonance phenomena, as will be discussed later.
\vspace{-0.5cm}

\begin{figure}[htb]
\rotatebox{0}{\includegraphics[width=140mm,height=60mm]{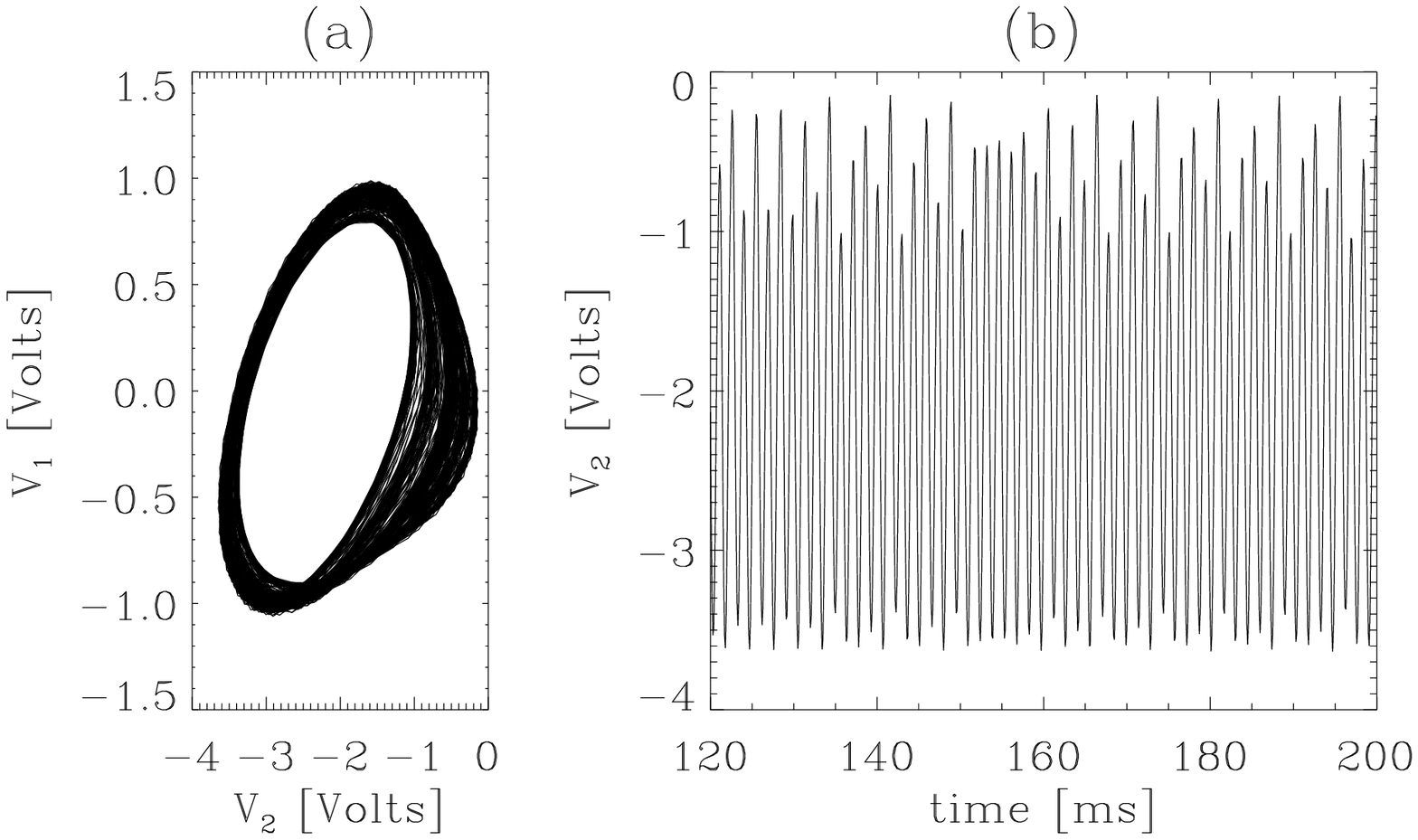}}
\label{fig2}
\end{figure}
\vspace{-0.5cm}

\begin{center} 
{\bf FIGURE 2} (a) Phase space representation of the attractor; (b) Time series of
$V_2$
\end{center}

To study the stochastic resonance effect we start by adding a small periodic
signal (a sinusoidal one) to the circuit in the branch of
the impedance $L$ as shown in figure 1. The amplitude of the signal is
sufficiently small ($\sim 20$ mV) that the system is not able to jump to
its mirror image attractor. The frequency of the signal is $\sim
2.1$ KHz, close to the mean frequency of the Chua attractor. A white
noise source is also added to the system through the same branch of
the periodic signal source. We have checked that in the absence of
the periodic signal, the system remains in the same attractor, and
does not jump to its mirror image, for small noise levels. Once the
noise is increased the system starts to jump from one attractor to
the other. This situation is depicted in Figure 3(a), where we plot
the time evolution of $V_2$ for three different noise levels, together
with the sinusoidal signal. In a typical stochastic resonance behavior,
e.g. when a system jumps between two potential wells with stable
minima, its is expected that an optimum noise level exists for which
the system jumps at every external signal period. In our case, however, we do not
move between two stable minima but between two chaotic attractors, each
with its own internal dynamics. When looking in more detail at the
jumping process between the attractors, we observe that after a jump
the system starts to move close to the center of the attractor and
not in the outer orbits. In this case it is unlikely that the system would jump to the other attractor until it reaches an outer orbit. Therefore, for the sinusoidal frequency we have chosen, we do
not expect a jump from one attractor to the other at every sinusoidal
period. This fact is also reflected in figure 3(a). The center panel
corresponds to the time trace of $V_2$ with the optimum noise level.
Although the resonance between the noise and the external periodic
signal is not clear, it becomes evident when plotting the Fourier
transform of the signal, $\mid S(\nu)\mid$ (power spectrum). In figure 3(b)
it can be  observed a maximum of the Signal-to-Noise Ratio (SNR) at
an intermediate noise level, in this case a value of $\sim 1.5$ V
[rms]. This fact is confirmed when plotting the SNR vs. the noise
level (see figure 4). A clear maximum appears indicating
an optimum response of the system to this intermediate noise
intensity.
\vspace{-1.2cm}

\begin{figure}[ht]
\begin{minipage}{0.5\textwidth}
\rotatebox{0}{\includegraphics[width=8.0cm,height=125mm ]{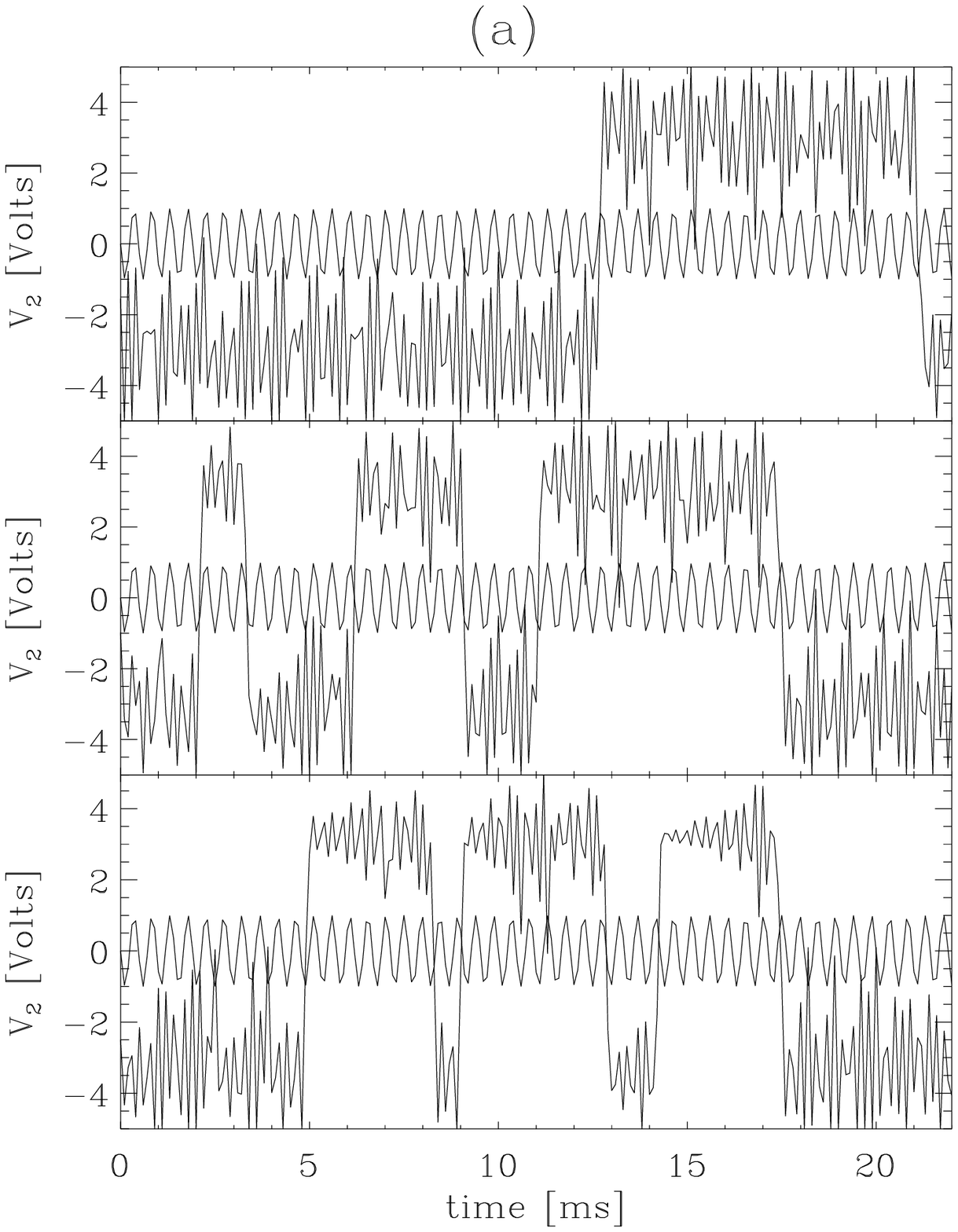}}
\end{minipage}\hfill
\begin{minipage}{0.5\textwidth}
\rotatebox{0}{\includegraphics[width=8.0cm,height=125mm]{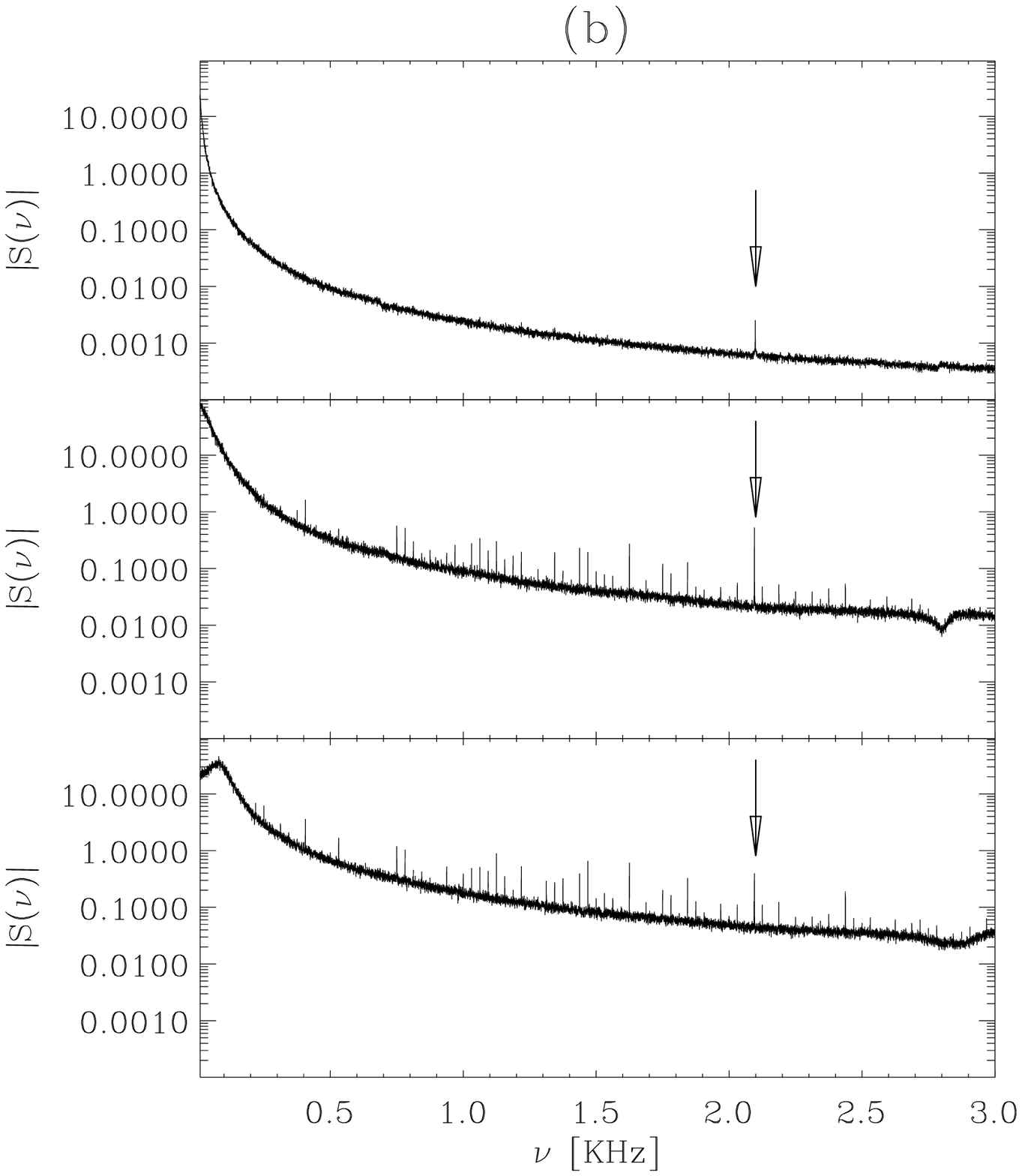}}
\end{minipage}
\label{fig3}
\end{figure}
\vspace*{-2.cm}

\noindent {\bf FIGURE 3} (a) Time evolution of $V_2$ for three different noise levels. From top
to bottom: 1V [rms], 1.5 V [rms] and 2.25 V [rms]. The sinusoidal
signal is added in the three panels with an arbitrary amplitude; (b)
Power spectrum for the same noise intensity as in panel (a). The arrow indicates the frequency of the sinusoidal signal.

\begin{figure}[htb]
\rotatebox{0}{\includegraphics[width=100mm,height=60mm]{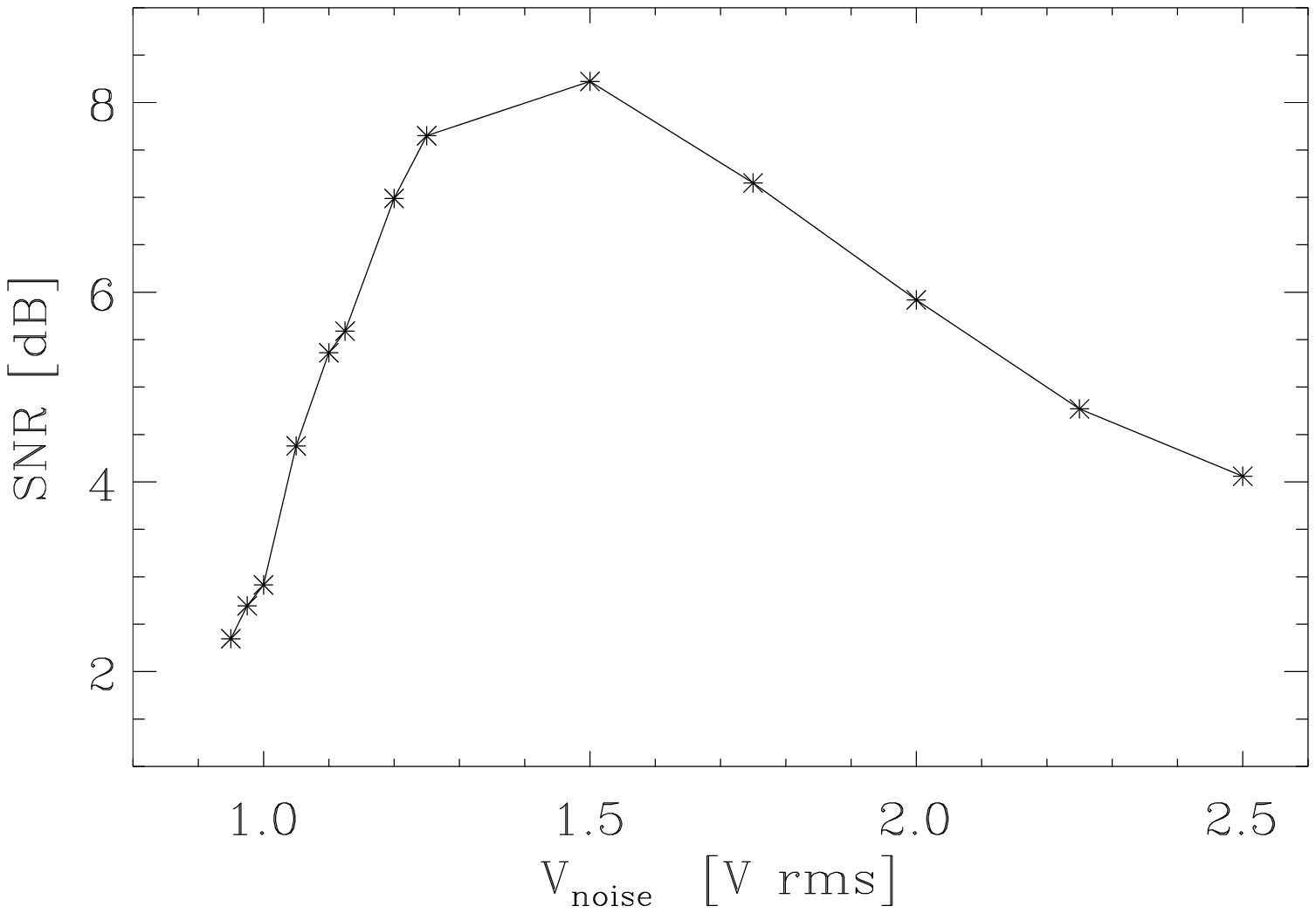}}
\label{fig4}
\end{figure}
\vspace*{-0.3cm}

\begin{center} 
{\bf FIGURE 4} Signal-to-noise ratio as a function of the noise
level.
\end{center}

The phenomenon of coherence resonance is characterized by an optimum response
of a system for an intermediate noise level without any other
forcing. In this sense it is not very different to the stochastic
resonance except for the absence of the sinusoidal source, and for
this reason it is sometimes called {\sl stochastic resonance without
external forcing}. We start our experiment in the same situation as we
started the previous one. Again for small noise levels the system is
not able to jump to its mirror image attractor. As we do not have a
periodic external forcing we do not expect any resonance in the SNR.
This means that we have to look for other indicators. To quantify the
situation we calculate the variance $\sigma$ of the residence time in
the attractors normalized to the mean value $\langle t \rangle$ and
the normalized autocorrelation function of the time series\cite{PK97}.  Three
typical time traces are shown in the figure 5(a) for the voltage
$V_2$ at low, intermediate and large noise levels. It can be seen
that for the intermediate noise level (the optimum one), the
regularity of the jumping process becomes evident. This is the first
indication that noise is inducing a more regular behavior in this
chaotic system. The autocorrelation functions depicted in figure
5(b) present the deepest minimum value at the same noise level,
indicating maximum anticorrelation between the two states. These
results reveal the existence of a constructive effect of the noise
that is capable to yield a maximum regularity in the process of
transition from one chaotic attractor to the other.
\vspace{-1.2cm}

\begin{figure}[htb]
\begin{minipage}{0.5\textwidth}
\rotatebox{0}{\includegraphics[width=7.0cm,height=100mm]{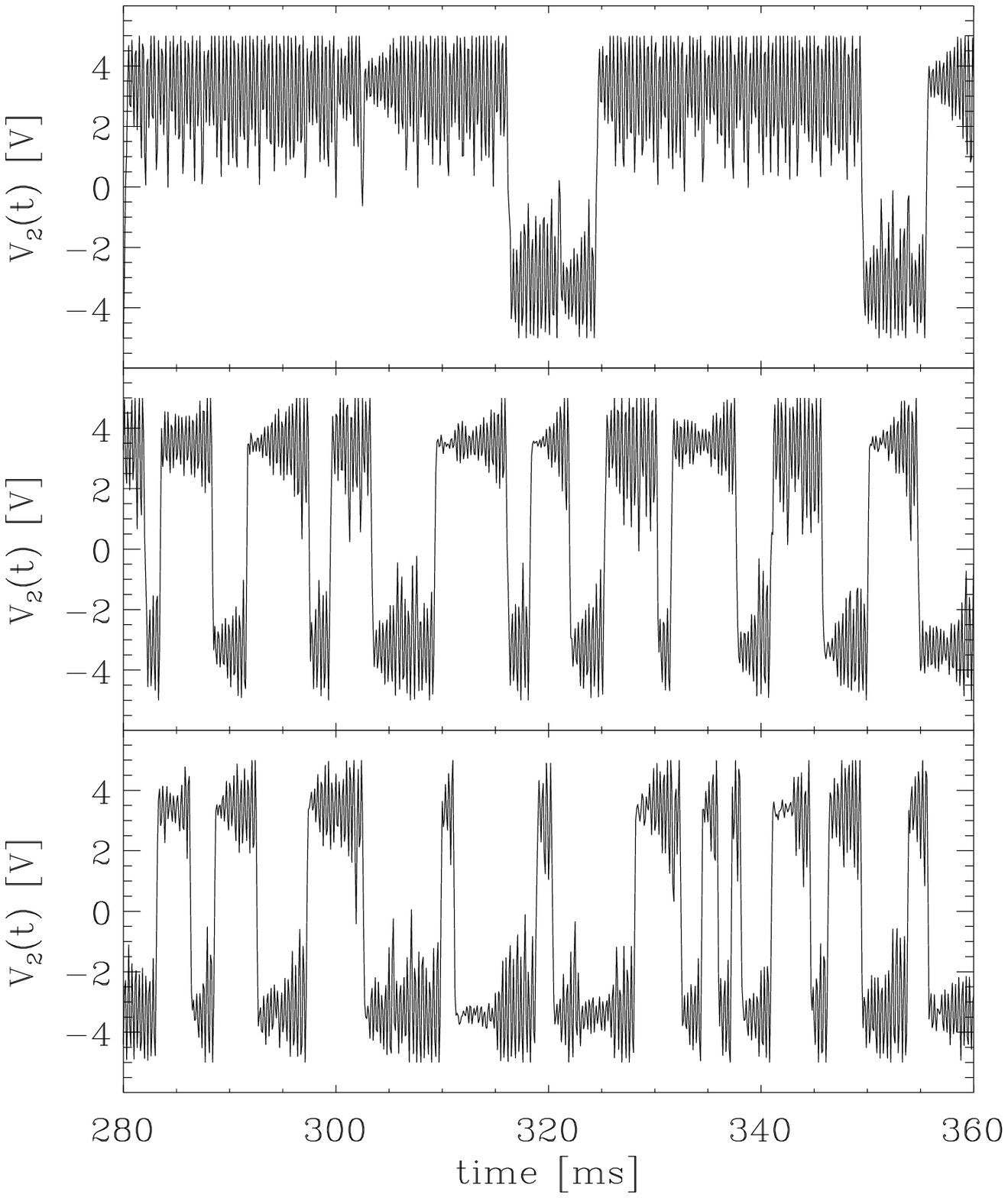}}
\end{minipage}\hfill
\begin{minipage}{0.5\textwidth}
\rotatebox{0}{\includegraphics[width=7.0cm,height=100mm]{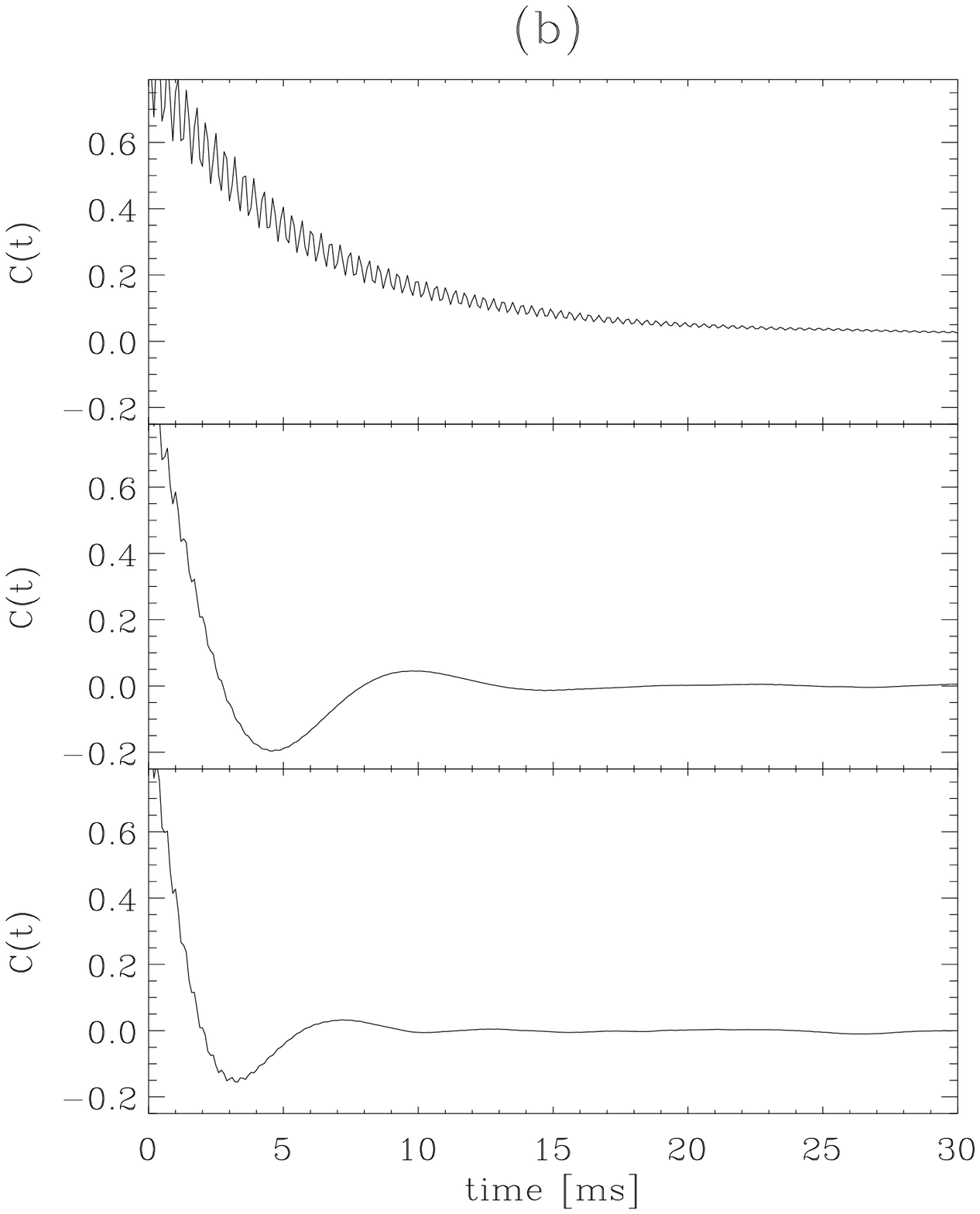}}
\end{minipage}
\label{fig5}
\end{figure}
\vspace*{-2.cm}

\noindent {\bf FIGURE 5} (a) Voltage time series for three different noise levels
(maximum values): 0.5 V[rms], 1.5 V[rms], optimum noise
level, and 2.5 V[rms]; (b) Autocorrelation function for the same
noise levels.
\vspace*{0.5cm}

Finally, in figure 6 we plot both the normalized variance of
the residence time and the minima of the autocorrelation functions vs.
the noise levels. Both quantities have a minima close to the noise
intensity $\sim 1.5$ V[rms].

In conclusion, we have experimentally shown that the main features
of both stochastic and coherence resonance are present in some
chaotic systems. We have done the demonstration using a chaotic Chua
circuit. In the case of the stochastic resonance we observed
a clear maximum for an intermediate noise level of the signal-to-noise ratio at the external
frequency. For the coherence resonance,  a quasi-periodic response
can be induced by just applying an  appropriate amount of noise to a
circuit which, otherwise, exhibits a  non-regular behavior. The
quality of the periodicity was quantified  by the variance of the
residence time and the minimum of the time  correlation function.

\begin{figure}[htb]
\rotatebox{0}{\includegraphics[width=100mm,height=80mm]{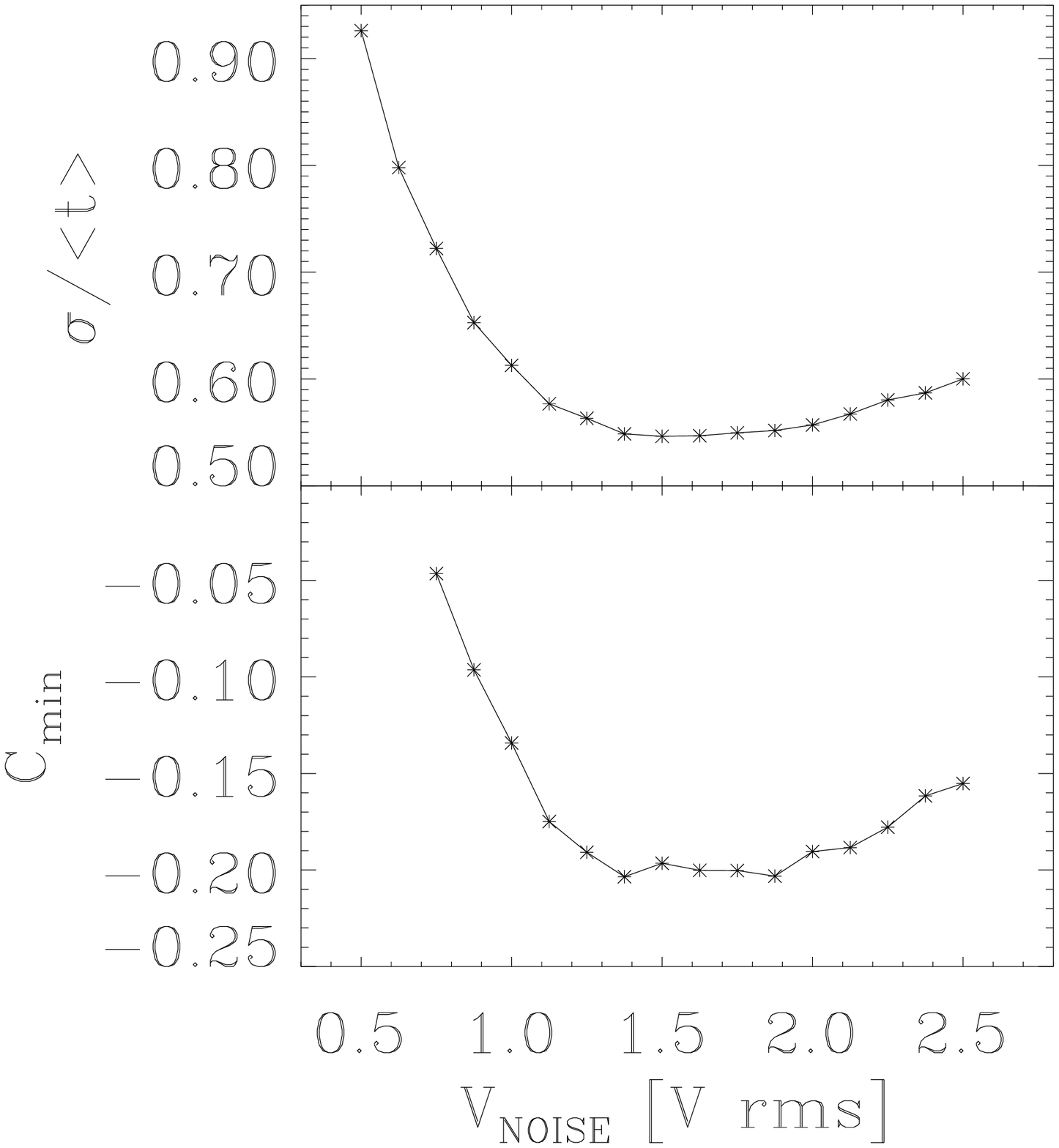}}
\label{fig6}
\end{figure}
\vspace*{-1.cm}

\noindent {\bf FIGURE 6}
(a) Variance of the residence time normalized by the mean
time $\sigma/\langle T \rangle$ and (b) minimum of the correlation
function $C_{min}$ as a function of the noise level
$V_{NOISE}$.

We thank financial support from DGESIC (Spain)
projects numbers PB97-0141-C02-01 and BFM2000-1108.
\doingARLO[\bibliographystyle{aipproc}]
          {\ifthenelse{\equal{\AIPcitestyleselect}{num}}
             {\bibliographystyle{arlonum}}
             {\bibliographystyle{arlobib}}
          }

\end{document}